\documentclass[a4paper,11pt]{article}
\usepackage{amssymb}

\input{tcilatex}
\usepackage{graphicx}
\usepackage{graphicx}
\begin{document}

\title{Energy Gradient Theory of Hydrodynamic Instability }
\author{Hua-Shu Dou \\
Fluid Mechanics Division\\
Department of Mechanical Engineering\\
National University \emph{of} Singapore\\
Singapore 119260\\
Email: mpedh@nus.edu.sg; huashudou@yahoo.com\\
**********\\
\emph{(This paper has been presented at }\\
\emph{The Third International Conference on }\\
\emph{Nonlinear Science, }\\
\emph{Singapore, 30 June -- 2 July, 2004,}\\
http://www.ddap3.nus.edu.sg\emph{)  }}
\date{}
\maketitle

\begin{abstract}
A new universal theory for flow instability and turbulent transition is
proposed in this study. Flow instability and turbulence transition have been
challenging subjects for fluid dynamics for a century. The critical
condition\ of turbulent transition from theory and experiments differs
largely from each other for Poiseuille flows. This enigma has not been
clarified so far owing to the difficulty of the problem. In this paper, a
new mechanism of flow instability and turbulence transition is presented for
parallel shear flows and the energy gradient theory of hydrodynamic
instability is proposed. It is stated that the total energy gradient in the
transverse direction and that in the streamwise direction of the main flow
dominate the disturbance amplification or decay. Thus, they determine the
critical condition of instability initiation and flow transition under given
initial disturbance. A new dimensionless parameter $K$ for characterizing
flow instability is proposed for wall bounded shear flows, which is
expressed as the ratio of the energy gradients in the two directions. It is
thought that flow instability should first occur at the position of $K_{\max
}$ which may be the most dangerous position. This speculation is confirmed
by Nishioka et al's experimental data. Comparison with experimental data for
plane Poiseuille flow and pipe Poiseuille flow indicates that the proposed
idea is really valid. It is found that the turbulence transition takes place
at a critical value of $K_{\max }$ of about $385$ for both plane Poiseuille
flow and pipe Poiseuille flow, below which no turbulence will occur
regardless the disturbance. More studies show that the theory is also valid
for plane Couette flows and Taylor-Couette flows between concentric rotating
cylinders. It is concluded that the energy gradient theory is a universal
theory for the flow instability and turbulent transition which is valid for
both pressure and shear driven flows in both parallel flow and rotating flow
configurations.
\end{abstract}


\newpage

\section{Introduction}

Understanding the mechanism of turbulence has been a great challenge for
over a century. Now, it is still very far from approaching a comprehensive
theory and the final resolution of the turbulent problem \cite{lumly}.
Reynolds (1883)[1] \cite{reynolds} did the first famous experiments on pipe
flow demonstrating the transition from laminar to turbulent flows. Since
then, various stability theories emerged during the past 120 years for this
phenomenon, but few are satisfactory in the explanation of the various flow
instabilities and the related complex flow phenomena.

The pipe Poiseuille flow (Hagen-Poiseuille) is linearly stable for all the
Reynolds number $Re$ by eigenvalue analysis \cite{landau} \cite{drazin} \cite%
{schmid01} \cite{trefethen93} \cite{gross00}. However, experiments showed \
that the flow would become turbulence if $Re$ ($=\rho UD/\mu $) exceeds a
value of $2000$. Experiments also showed that disturbances in a laminar flow
could be carefully avoided or considerably reduced, the onset of turbulence
was delayed to Reynolds numbers up to $Re=O(10^{5})$ \cite{trefethen93} \cite%
{gross00}. For $Re$ above $2000$, the characteristic of turbulence
transition depends on the disturbance amplitude and frequency, below which
transition from laminar to turbulent state does not occur regardless of the
initial disturbance amplitude \cite{wygnanski} \cite{darby}. Thus, it is
clear that the transition from laminar to turbulence for $Re>2000$ is
dominated by the behaviours of mean flow and disturbance. Only the combined
effect of the two factors reaches the critical condition, could the
transition occur for $Re>2000$. These are summarized in Table 1.

Linear stability analysis of plane parallel flows gives critical Reynolds
number $Re$ ($=\rho V_{0}h/\mu $) of $5772$ for plane Poiseuille flow, while
experiments show that transition to turbulence occurs at Reynolds number of
order $1000$ \cite{orszag71} \cite{landau} \cite{trefethen93} \cite{gross00}%
, even though the laminar flow could also kept to $Re=O(10^{5})$ \cite%
{Nishioak75}. One resolution of these paradoxes is that the domain of
attraction for the laminar state shrinks for large Re (as $\func{Re}^{\gamma
}$ say, with $\gamma <0),$ so that small but finite perturbations lead to
transition \cite{trefethen93} \cite{chapamn2002}. Grossmann \cite{gross00}
commented that this discrepancy demonstrates that nature of the
onset-of-turbulence mechanism in this flow must be different from an
eigenvalue instability. Orszag and Patera \cite{orszag80} remarked that the
mechanism of transition is not properly represented by parallel-flow linear
stability analysis. They proposed a linear three-dimensional mechanism to
predict the transitional Reynolds number. Some nonlinear stability theories
have been proposed, for example, in \cite{stuart}. However, these theories
do not seem to offer a good agreement with the experimental data.

Energy method was also used in the study of flow instabilities \cite%
{LinCC1955} \cite{Betchov} \cite{joseph76} \cite{drazin} \cite{hinze} \cite%
{schmid01}. In energy method, one observes the rate of increasing of
disturbance energy to study the instability of the \emph{flow system}. The
critical condition is determined by the maximum Reynolds number at which the
disturbance energy in the system monotonically decreases. In the flow
system, it is considered that turbulence shear stress interacts with the
velocity gradient\ and the disturbance gets energy from mean flow in such a
way. Thus, the disturbance is amplified and the instability occurs with the
energy increasing of disturbance. Therefore, it is recognized that it is the
basic state vorticity leading to instability. The energy method could not
get agreement with the experiments either \cite{Betchov} \cite{drazin} \cite%
{schmid01}. In recent years, various transition scenarios have been proposed 
\cite{trefethen93} \cite{gross00} \cite{Waleffe} \cite{baggett} \cite%
{Zikanov} \cite{reddy1998} \cite{chapamn2002} \cite{Meseguer} for the
subcritical transition. Although we can get a better understanding of the
transition process from these scenarios, the mechanism is still not fully
understood and the agreement with the experimental data is still not
satisfied.

Generally, the transition from laminar flow to turbulent flow is not
generated suddenly in the entire flow field but it first starts from
somewhere in the flow field and then spreads out gradually from this
position. As is well known in solid mechanics, the damage of a metal
component generally starts from some area such as manufacturing fault,
crack, stress concentration, or fatigue position, etc. In fluid mechanics,
we consider that the breaking down of a steady laminar flow should also
start from a most dangerous position first. The consequent questions are:
(a) \emph{Where is this most dangerous position for Poiseuille flow? }(b) 
\emph{What is the mechanism and the dominant factor for this phenomenon?}
(c) \emph{What parameter should be used to characterize this position?}
These questions are our concern. Finding the solution of these problems is
important to the understanding of the phenomenon and the estimation of flow
transition. Because the turbulence transition is generally resulted in by
flow instability \cite{LinCC1955}, we think that the critical condition of
transition should be determined by the position where the \emph{flow
instability} first takes place. If the mechanism of flow instability is
sought out and the most dangerous position is found, the critical condition
of transition could be determined.

\begin{table}[tbp] \centering%
\begin{tabular}{|l|l|l|}
\hline
Base flow & Disturbance & Resulting flow state \\ \hline
$Re<2000$ & No matter how large. & Disturbance decay; flow keeps laminar. \\ 
\hline
$Re\thicksim 10^{5}$ & Kept low. & Flow keeps laminar. \\ \hline
$Re\geq 2000$ & Enough large, depend on $Re$. & Transition occurs. \\ \hline
\end{tabular}%
\caption{Characteristic of transition of Hagen-Poiseuille flow (pipe
Poiseuille flow).}%
\end{table}%

In this study, we explore the critical condition of main flow for
instability and turbulence transition, and not deal with the detailed
process of disturbance amplification. The energy gradient theory is proposed
to explain the mechanism of flow instability and turbulence transition for
parallel flows. A new dimensionless parameter for characterizing the
critical condition of flow instability is proposed. Comparison with
experimental data for plane Poiseuille flow and pipe Poiseuille flow at
subcritical transition indicates that the proposed idea is really valid.

\section{Proposed Mechanism of Flow Instability}

\FRAME{ftbpFU}{3.4826in}{3.0952in}{0pt}{\Qcb{Velocity distribution variation
with the increased Reynolds number for given fluid and geometry in plane
Poiseuille flows. $Re=\protect\rho UL/\protect\mu ,$ $L=2h,$ where $h$ is
the half-width of the channel.}}{}{parab1.ps}{\raisebox{-3.0952in}{\includegraphics[height=3.0952in]{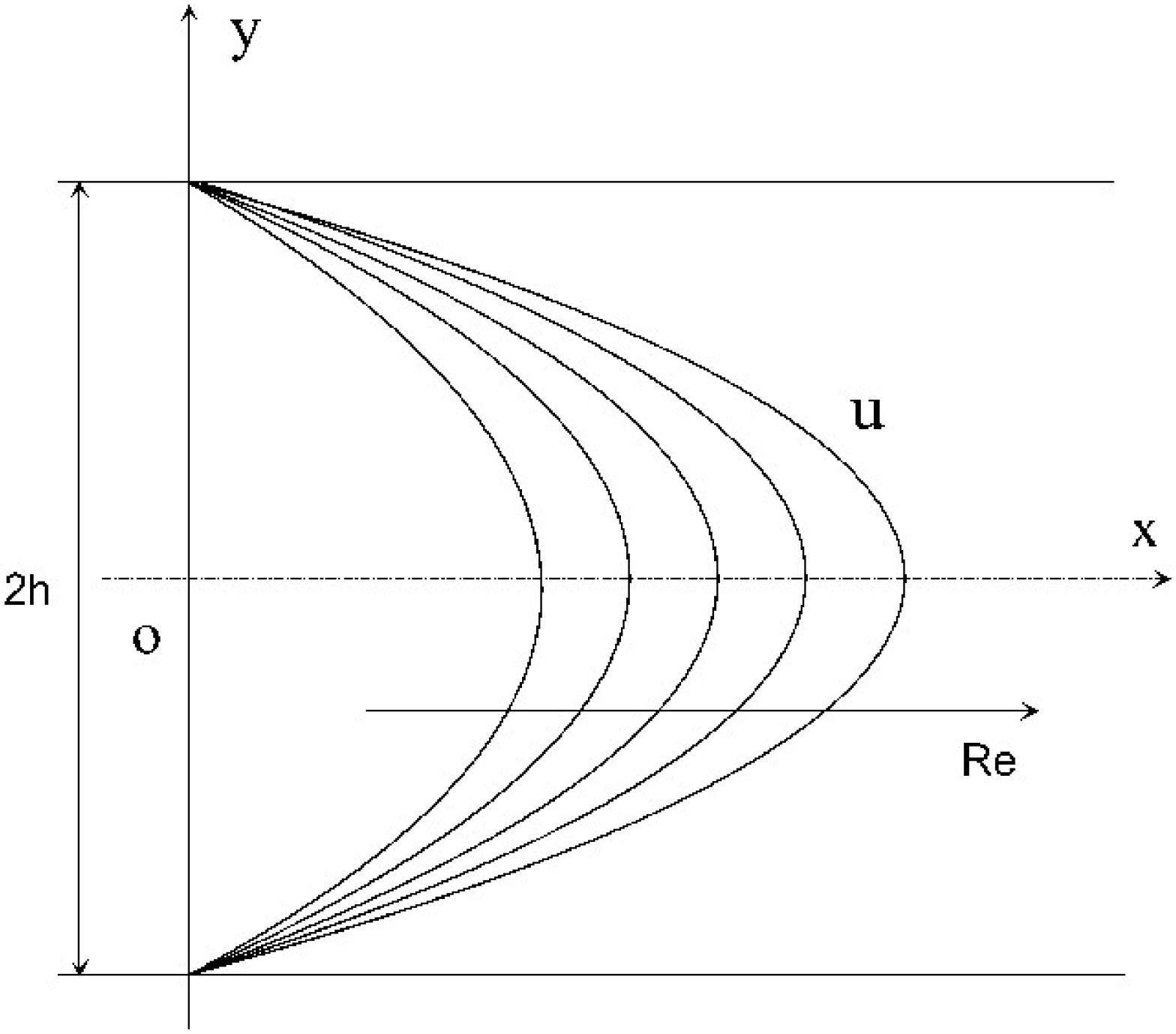}}}

\bigskip \FRAME{ftbpFU}{4.2203in}{3.0943in}{0pt}{\Qcb{Velocity profiles for
laminar and turbulent flows.}}{}{parabb1.ps}{\raisebox{-3.0943in}{\includegraphics[height=3.0943in]{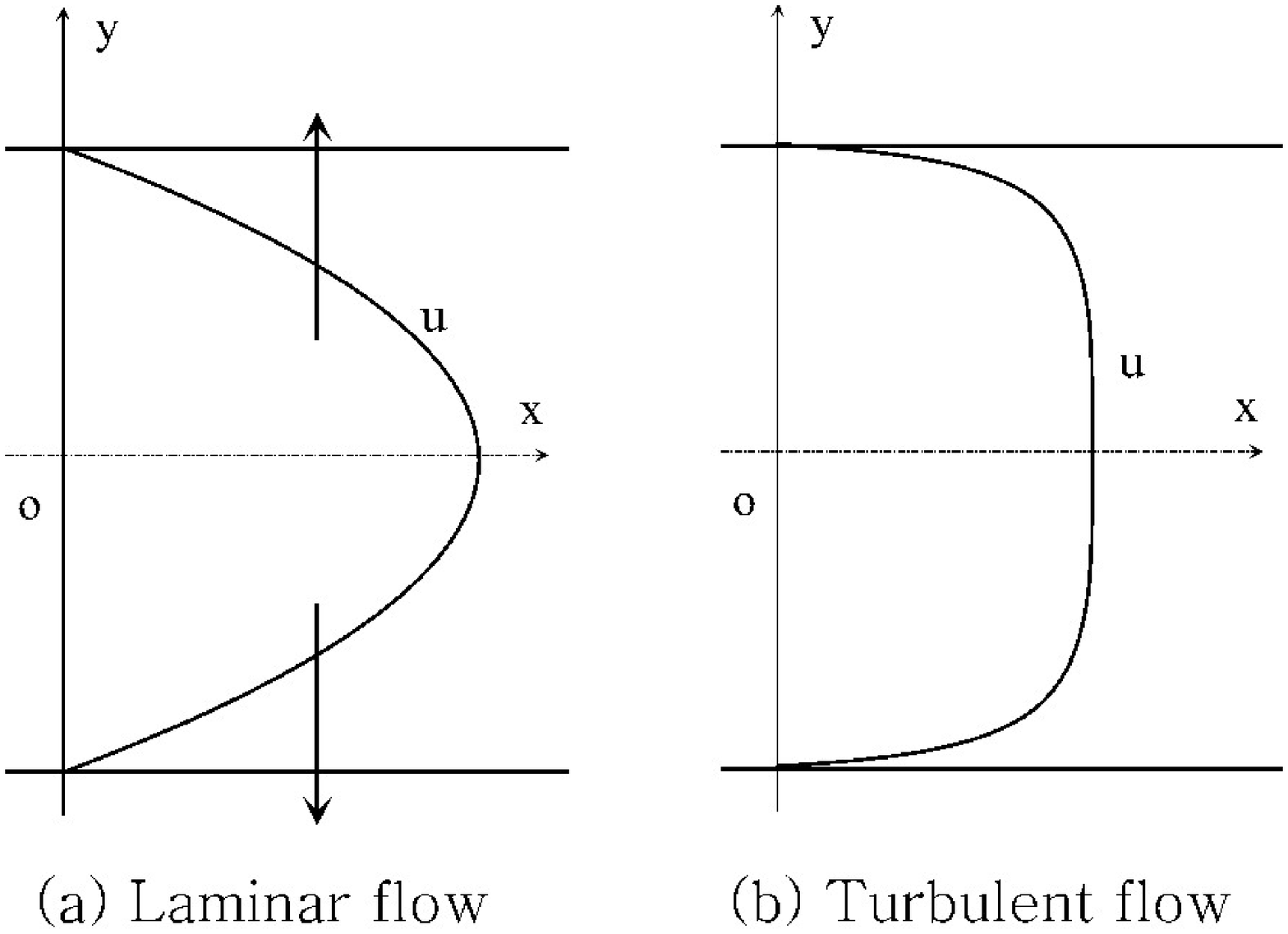}}}

The plane Poiseuille flow in a channel is shown in Fig.1. For the given flow
geometry and fluid property, with the increasing mean velocity $U$, the flow
may transit to turbulence if the $Re$ exceeds a critical value (under
certain disturbance). The velocity profiles for laminar and turbulent flows
are shown respectively in Fig.2. It can be imagined that there is a
\textquotedblleft driving factor\textquotedblright\ which pulls the laminar
velocity profile outward toward the walls when the transition takes place.
What should be such a driving factor? From a large amount of observations,
it is thought that\emph{\ }the increase of the gradient of fluid kinematic
energy in transverse direction, $\frac{\partial }{\partial y}(\frac{1}{2}%
V^{2})=V\frac{\partial V}{\partial y}$, may form a \textquotedblleft \emph{%
driving force}\textquotedblright\ to cause the increase of flow disturbance
for given flow condition, while the gradient of the viscous friction force
may resist or absorb the disturbance. Here, $V$ is the magnitude of local
velocity. The stability of the flow depends upon the effects of these two
roles. With the increasing of mean velocity $U$ for parallel flows, the
energy gradient in the transverse direction increases. If this energy
gradient is large enough it will lead to a disturbance amplification of the
flow. The viscosity friction caused by shear stress would stabilize the flow
by absorbing the velocity fluctuation. When the energy gradient in
transverse direction reaches beyond a critical value, the laminar flow could
not balance this disturbance and flow instability might be excited. Finally,
the turbulence flow would be triggered when the transverse energy gradient
continuously keeps large enough with the flow forward. The energy gradient
in the transverse direction makes the exchange of energy between the fluid
layers and sustains the turbulence. Therefore, it is proposed that \emph{the
necessary condition for the turbulence transition is that there is an energy
gradient in the transverse direction of the main flow.}

Now, we prove this necessary condition is correct at least for parallel
flows. If the gravitational energy is neglected, the total energy gradient
in transverse direction is $\frac{\partial }{\partial y}\left( p+\frac{1}{2}%
\rho V^{2}\right) $. For parallel flows, $\frac{\partial p}{\partial y}=0$
and $V=u$. If this energy gradient is zero, $\frac{\partial }{\partial y}(%
\frac{1}{2}V^{2})=V\frac{\partial V}{\partial y}=0$, there must be $\frac{%
\partial V}{\partial y}=0$ due to $V\neq 0$. Thus, the rate of increase of
disturbance energy will be negative due to viscous dissipation because the
disturbance could not obtain energy from the base flow owing to zero
velocity gradient \cite{Betchov} \cite{drazin} \cite{schmid01}. Therefore,
the disturbance must decay at this case. In such way, it is proved that the
energy gradient in the transverse direction is a necessary condition for the
flow transition.

In addition, when there is a pressure gradient in the normal direction to
the flow direction, this pressure gradient could also result in a flow
instability even the Reynolds number is low. Both centrifugal and Coriolis
instabilities are those caused by pressure gradients. Elastic instability is
also that produced by the transversal pressure gradient \cite{shaqfeh96} 
\cite{dou02a}. The mechanism of instability should take into account of the
effect of the variation of cross-streamline pressure for these cases, which
may lead to flow instability or accelerates the instability initiation.{\ }%
In some cases of incompressible flows such as stratified flows, the
gravitational energy should be taken into account.

For given flow geometry and fluid, it is proposed that the flow stability
condition can be expressed as, 
\begin{equation}
\frac{\partial }{\partial y}\left( \rho g_{y}y+p+\frac{1}{2}\rho
V^{2}\right) <C\text{,}  \label{E1}
\end{equation}%
where $g_{y}$ is the component of gravity acceleration in $y$ direction, and 
$C$ is a constant\ which is related to fluid property and geometry. The $x$
axis is along the flow direction and the $y$ axis is along the transverse
direction. In this study, we first show that the proposed idea is really
correct for Poiseuille flows.

\section{Formulation and Theory Description}

The conservation of momentum for an incompressible Newtonian fluid is
(neglecting gravity force): {\ 
\begin{equation}
\rho (\frac{\partial \mathbf{V}}{\partial t}+\mathbf{V}\cdot \nabla \mathbf{V%
})=-\nabla p+\mu \nabla ^{2}\mathbf{V}\text{.}  \label{E3}
\end{equation}%
Using the identity, 
\begin{equation}
\mathbf{V}\cdot \nabla \mathbf{V}=\frac{1}{2}\nabla \left( \mathbf{V}\cdot 
\mathbf{V}\right) -\mathbf{V}\times \nabla \times \mathbf{V}\text{,}
\label{E4}
\end{equation}%
equation (\ref{E3}) can be rearranged as, 
\begin{equation}
\rho \frac{\partial \mathbf{V}}{\partial t}+\nabla (p+\frac{1}{2}\rho
V^{2})=\mu \nabla ^{2}\mathbf{V}+\rho \mathbf{(V}\times \nabla \times 
\mathbf{V)}\text{,}  \label{E5}
\end{equation}%
}where $\rho $ is the fluid density, $t$ the time, $\mathbf{V}$ the velocity
vector, $p$ the hydrodynamic pressure, $\mu $ the dynamic viscosity of the
fluid. If the viscous force is zero, the above equation becomes the Lamb
form of momentum equation. This equation can be found in most text books.
For incompressible flow, the total pressure represents the total energy in
Eq. (\ref{E5}). Actually, the energy equation has long been used for
stability analysis as previously mentioned \cite{hinze} \cite{drazin} \cite%
{schmid01} \cite{joseph76} \cite{Betchov}.

In previous sections, it is proposed that the instability of viscous flows
depends on the relative magnitude of the energy gradient in transverse
direction and the viscous friction term. A larger energy gradient in
transverse direction tries to lead to amplification of a disturbance, and a
large shear stress gradient in streamwise direction tends to absorb this
disturbance and to keep the original laminar flow state. The transition\ of
turbulence depends on the relative magnitude of the two roles of energy
gradient amplification and viscous friction damping under given disturbance.
We propose the parameter for characterizing the relative role of these
effects below.

Let $d\mathbf{s}$ represent the differential length along a streamline in a
Cartesian coordinate system,{\ 
\begin{equation}
d\mathbf{s}=d\mathbf{x}+d\mathbf{y}\text{.}
\end{equation}%
}

With dot multiplying Eq. (\ref{E5}) by $d\mathbf{s}$, we obtain,{\ 
\begin{equation}
\rho \frac{\partial \mathbf{V}}{\partial t}\cdot d\mathbf{s}+\nabla (p+\frac{%
1}{2}\rho V^{2})\cdot d\mathbf{s}=\mu \nabla ^{2}\mathbf{V}\cdot d\mathbf{s}%
+\rho \mathbf{(V}\times \nabla \times \mathbf{V)}\cdot d\mathbf{s}\text{.}
\label{E6}
\end{equation}%
}

Since $\mathbf{(V}\times \nabla \times \mathbf{V)}\cdot d\mathbf{s=}0$ along
the streamline, for steady flows, we obtain the energy gradient along the
streamline,{\ 
\begin{equation}
\partial (p+\frac{1}{2}\rho V^{2})/\partial s=\mu \nabla ^{2}\mathbf{V}\cdot 
\frac{d\mathbf{s}}{\left\vert d\mathbf{s}\right\vert }\mathbf{=}(\mu \nabla
^{2}\mathbf{V)}_{s}\text{.}  \label{E7}
\end{equation}%
This equation shows that t}he total energy gradient along the streamwise
direction equals to the viscous term $(\mu \nabla ^{2}\mathbf{V)}_{s}$. For
pressure driving flows, $(\mu \nabla ^{2}\mathbf{V)}_{s}$ represents the
energy loss due to friction. It is obvious that the total energy decreases
along the streamwise direction due to viscous friction loss. The energy
gradient along the transverse direction is,{\ 
\begin{equation}
\partial (p+\frac{1}{2}\rho V^{2})/\partial n=\frac{\partial p}{\partial n}%
+\rho V\frac{\partial V}{\partial n}\text{.}  \label{E8}
\end{equation}%
It can be seen that the energy gradient at transverse direction depends on
the velocity gradient and the velocity magnitude as well as the transversal
pressure gradient. }

The relative magnitude of the energy gradients in the two directions can be
expressed by a \emph{new dimensionless parameter}, $K,$ the ratio of the
energy gradient in the transverse direction to that in the streamwise
direction, {\ 
\begin{equation}
K=\frac{\partial E/\partial n}{\partial E/\partial s}=\frac{\partial (p+%
\frac{1}{2}\rho V^{2})/\partial n}{\partial (p+\frac{1}{2}\rho
V^{2})/\partial s}=\frac{\partial p/\partial n+\rho V(\partial V/\partial n)%
}{(\mu \nabla ^{2}\mathbf{V)}_{s}}\text{,}  \label{KNum}
\end{equation}%
where }$E=p+\frac{1}{2}\rho V^{2}$ expresses the total energy, $n$ denotes
the direction normal to the streamwise direction, and $s$ denotes the
streamwise direction.

It is noticed that the parameter $K$ is a field variable and it{\ represents
the direction of the vector of local total energy gradient. It can be seen
that when }$K$ is small, the role of the viscous term in Eq.(9) is large and
the flow tends to be stable. When $K$ is large, the role of numerator in
Eq.(9) is large and the flow tends to be unstable. {For given flow field,
there is a maximum of }$K$ in the domain which represents the most possible
unstable location. Therefore, in the area of high value of $K$, the flow
tends to be more unstable than that in the area of low value of $K$. The
first instability should be initiated by the maximum of $K$, $K_{max}$, in
the flow field for given disturbance. In other words, the position of
maximum of $K$ is the most dangerous position. The magnitude of $K_{max}$ in
the flow field symbolizes the extent of the flow approaching the
instability. Especially, if $K_{max}=\infty $, the flow is certainly
potential to be unstable under some disturbance. For given flow disturbance,
there is a critical value of $K_{max}$ over which the flow becomes unstable.
In particular, \emph{corresponding to the subcritical transition in wall
bounded parallel flows, the }$K_{max}$\emph{\ reaches its critical value,
below which no transition occurs regardless of the disturbance amplitude. }\
For unidirectional parallel flows, this critical value of $K_{max}$ should
be a constant regardless of the fluid property and the magnitude of the
geometrical parameter. Now, owing to the complexity of the flow, it is
difficult to predict this critical value by theory. Nevertheless, it can be
determined by the experimental data for given flows.

For parallel flows, the coordinates $s-n$ becomes the $x-y$ coordinates and
the pressure gradient $\partial p/\partial n=\partial p/\partial y=0$. Thus,
using the global approximation, $\partial (\frac{1}{2}\rho V^{2})/\partial
y\backsim \rho U^{2}/L$, $\left( \mu \nabla ^{2}\mathbf{V}\right)
_{s}\backsim \partial ^{2}u/\partial y^{2}\backsim U/L^{2}$, we obtain $%
K\backsim $ $\rho UL/\mu =Re,$ where $L$ is a characteristic length in the
transverse direction. Thus, the parameter $K$ is equivalent to the Reynolds
number in the global sense for parallel flows.

The negative energy gradient in the streamwise direction plays a part of
resisting on or absorbing the disturbance. Therefore, it has a stable role
to the flow. The energy gradient in the transverse direction has a role of
amplifying disturbance. Therefore, it has a unstable role to the flow. Here,
it is not to say that a high energy gradient in transverse direction
necessarily leads to instability, but it has a \emph{potential for
instability}. Whether instability occurs also depends on the magnitude of
the disturbance. From this discussion, it is easily understood that the
disturbance amplitude required for instability is small for high energy
gradient in transverse direction (high Re) \cite{darby}.

\FRAME{ftbpFU}{5.5002in}{3.0952in}{0pt}{\Qcb{Schematic of energy gradient
and energy angle for plane Poiseuille flows. (a) Energy angle increases with
the Reynolds number; (b) Definition of the energy angle.}}{}{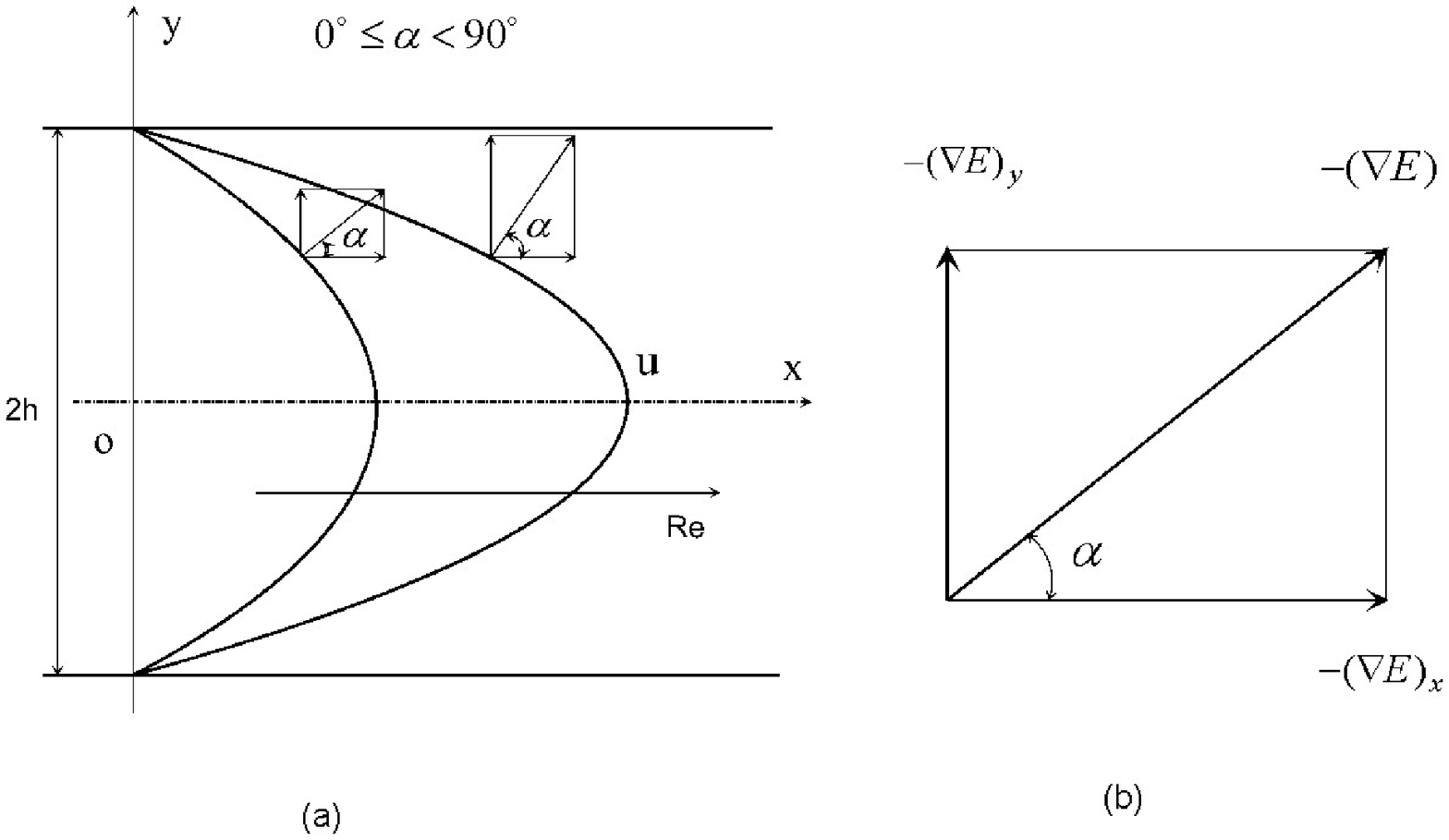}{%
\raisebox{-3.0952in}{\includegraphics[height=3.0952in]{angle3a0.ps}}}

\FRAME{ftbpFU}{2.4517in}{3.0943in}{0pt}{\Qcb{Schematic of the direction of
the total energy gradient and energy angle for flow with an inflection point
at which the energy angle equals 90 degree.}}{}{center2a.ps}{\special%
{language "Scientific Word";type "GRAPHIC";maintain-aspect-ratio
TRUE;display "USEDEF";valid_file "F";width 2.4517in;height 3.0943in;depth
0pt;original-width 4.2921in;original-height 5.431in;cropleft "0";croptop
"1";cropright "1";cropbottom "0";filename '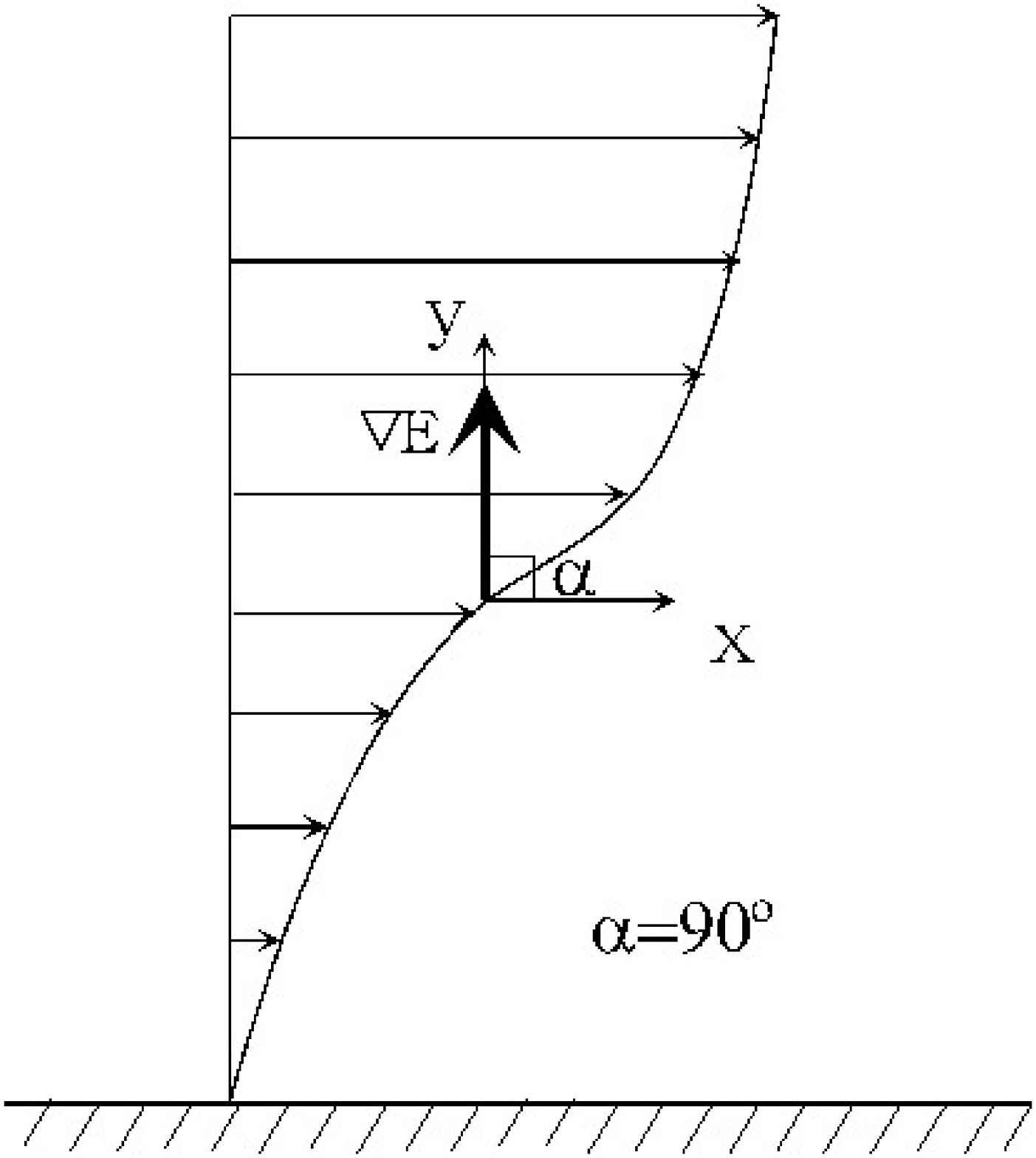';file-properties
"XNPEU";}}

{\ The value of \textquotedblleft }$\arctan K$\textquotedblright {\
expresses the angle between the direction of the total energy gradient and
the streamwise direction. We write that,}

\begin{equation}
\alpha =\arctan K\text{.}
\end{equation}%
{In this paper, the angle }$\alpha ${\ is named as \textquotedblleft \emph{%
energy angle},\textquotedblright\ as shown in Fig.3. }The value of the
energy angle (its absolute value) can also be used to express the extent of
the flow near the instability occurrence. Fig.3 and Fig.4 show the schematic
of the energy angle for some flows. There is a critical value of energy
angle, $\alpha _{c},$ corresponding to the critical value of $K_{max}$. When 
$\alpha >\alpha _{c}$, \ the flow becomes unstable. For Poiseuille flows, $%
0^{\circ }\leq \alpha <90^{\circ }$ and the stability depends on the
magnitudes of the energy angle and the disturbance. For parallel flow with a
velocity inflection, $\alpha =90^{\circ }$ ($K_{max}=\infty $) at the
inflection point and the flow is therefore unstable. This is for the first
time to theoretically prove that \emph{viscous flow with a velocity
inflection is unstable}. Inversely, as discussed before, viscous flow
without velocity inflection may not be necessarily stable, depending on the
flow conditions (as shown in table 1). These results are correct at least
for \emph{pressure driving flows}.

Fig.4 is a best explanation of inflectional instability for viscous flows.
According to Eq.(\ref{KNum}), the physics of the criterion presented in this
paper is easily understood. The energy gradient in the transverse direction
tries to amplify a small disturbance, while the energy loss due to friction
in the streamwise direction plays a damping part to the disturbance. The
parameter $K$ represents the relative magnitude of disturbance amplification
due to energy gradient and disturbance damping of viscous loss. When there
is no inflection point in the velocity distribution, the amplification or
decay of disturbance depends on the two roles above, i.e., the value of $K.$
When there is an inflection point in the velocity distribution, the viscous
term vanishes, and while the transversal energy gradient still exists at the
position of inflection point. Thus, even a small disturbance must be
amplified by the energy gradient at this location. Therefore, the flow will
be unstable. Rayleigh (1880) \cite{rayleigh} only proved that inviscid flow
with an inflection point is unstable, while we demonstrate here that viscous
flow with an inflection point is unstable. The Rayleigh's criterion was
derived from mathematics using the inviscid theory, but its physical meaning
is still not clear.

It is well known that viscosity plays dual role on the flow instability and
disturbance amplification \cite{drazin} \cite{hinze} \cite{schmid01} \cite%
{LinCC1955}. It can be seen from equation (\ref{KNum}) that the higher the
viscosity, the larger the viscous friction loss. Thus, the flow is more
stable. If the values of the vorticity and the streamwise velocity are high,
the transversal energy gradient will be high. Thus, the flow is more
unstable. From these discussion, it is known that \emph{viscosity mainly
plays a stable role to the initiation of flow instability at subcritical
transition} by affecting the base flow through the viscous friction of
streamwise velocity. This is consistent with criterion of the Reynolds
number.

Reynolds number represents the ratio of convective inertia force to the
viscous force in the Navier-Stokes equations as a dimensionless parameter.
However, the magnitude of the Reynolds number is only a global indication of
the flow states and a rough expression for the transition condition. At same 
$Re$ number, the behaviour of flow instability may be different due to the
different combination of the magnitude of viscosity with other parameters
when $Re$ is larger than an indifference Reynolds number, as shown by the
chart of the solution of the Orr-Sommerfeld equation \cite{hinze} \cite%
{white} \cite{schlichting}. The role of viscosity is complicated with the
variations of flow parameters and flow conditions. The generation of
turbulence is not simply caused by increasing the $Re$, but actually by
increasing the value of $K$. When $Re$ increases, it inevitably leads to the
increase of $K$ in the flow field. The magnitude of local disturbance for a
fixed point depends not only on the apparent $Re$ number, but also on the
local flow conditions. For steady Poiseuille flow, the convective inertia
term is zero and the flow becomes turbulent at high $Re$. We can see that
the occurrence of turbulence is not due to the convective inertia term in
this case. In Poiseuille flow, the local Reynolds number is high in the core
area along the centerline, but the degree of turbulence is low. In the area
near the wall, the local Reynolds number is low, but the degree of
turbulence is high. In uniform flows, high Reynolds number may not
necessarily leads to turbulence. In summary, the Reynolds number is only a
global parameter, and the $K$ is a local parameter which represents the
local behaviour of the flow and best reflects the role of energy gradient.

\section{Analysis on Poiseuille Flows}

\subsection{Plane Poiseuille Flow}

\FRAME{ftbpFU}{4.8222in}{3.1436in}{0pt}{\Qcb{Kinematic energy gradient and
viscous force term versus the axial pressure gradient for plane Poiseuille
flow for any position in the flow field. $\left\vert dp/dx\right\vert
\propto U\propto Re$ for given fluid and geometry.}}{}{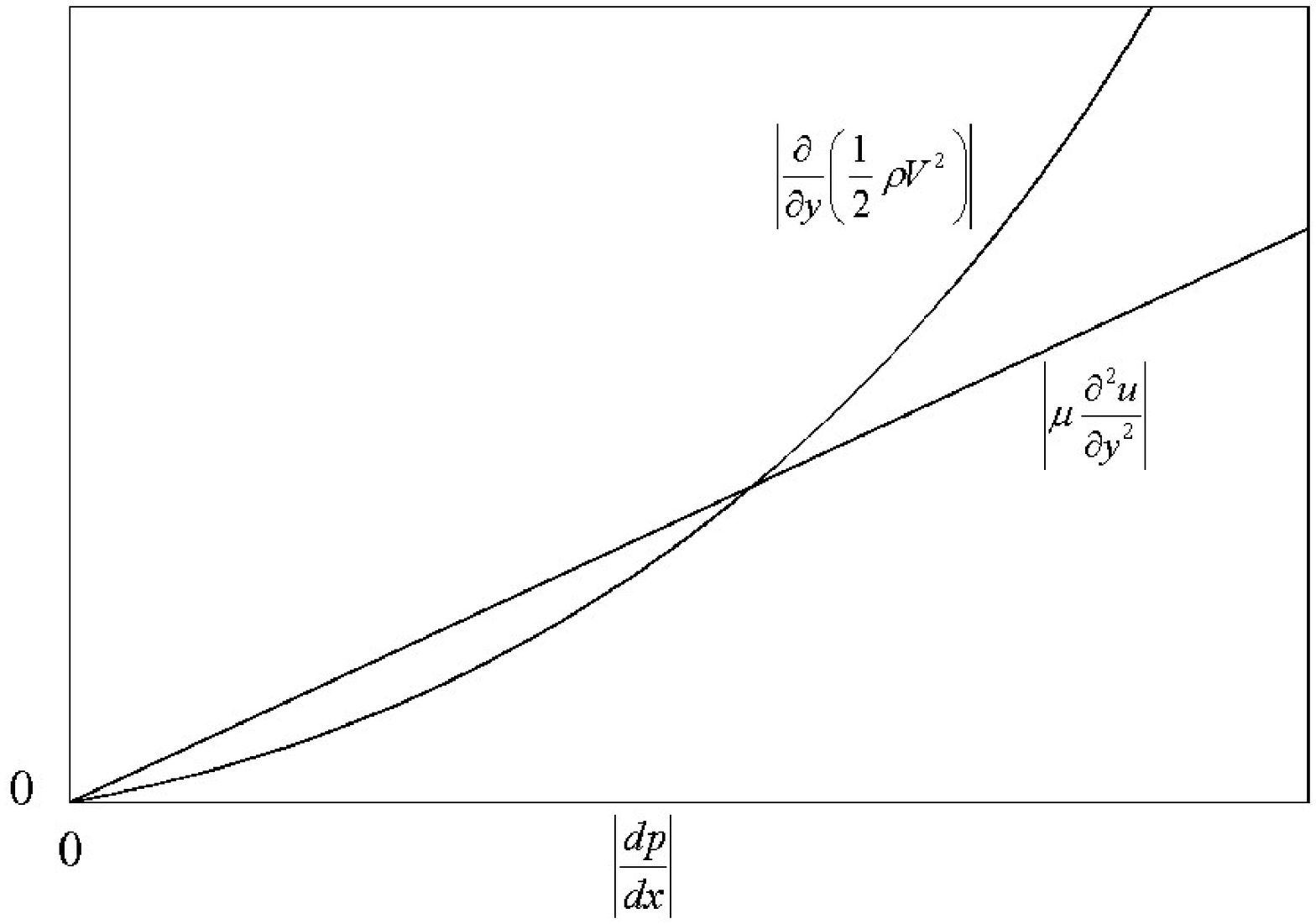}{\special%
{language "Scientific Word";type "GRAPHIC";maintain-aspect-ratio
TRUE;display "USEDEF";valid_file "F";width 4.8222in;height 3.1436in;depth
0pt;original-width 9.9895in;original-height 6.4895in;cropleft "0";croptop
"1";cropright "1";cropbottom "0";filename 'px1.ps';file-properties "XNPEU";}%
}

The instability in Poiseuille flows with increasing mean velocity $U$ can be
demonstrated as below for a given fluid and flow geometry by using the
energy gradient concept.

For 2D Poiseuille flow, the momentum equation is written as, {\ 
\begin{equation}
0=-\rho \frac{\partial p}{\partial x}+\mu \frac{\partial ^{2}u}{\partial
y^{2}}\text{,}  \label{P1}
\end{equation}%
showing that viscous force term is proportional to the streamwise pressure
gradient. The integration to above equation gives, 
\begin{equation}
u=u_{0}\left( 1-\frac{y^{2}}{h^{2}}\right) \text{,}
\end{equation}%
where }$u_{0}=-\frac{1}{2\mu }h^{2}\frac{\partial p}{\partial x}$, is the
centerline velocity, and $h$ is the channel half width.

{Although the energy gradient is not explicitly shown in above equation, it
can be expressed as below for any position in the flow field }(noticing $v=0$%
){, 
\begin{equation}
\frac{\partial }{\partial y}\left( \frac{1}{2}\rho V^{2}\right) =-\frac{\rho 
}{2\mu ^{2}}h^{2}y\left( 1-\frac{y^{2}}{h^{2}}\right) \left( \frac{\partial p%
}{\partial x}\right) ^{2}\text{.}  \label{P2}
\end{equation}%
Thus, for any position in the flow field,} {\ 
\begin{equation}
\left\vert \frac{\partial }{\partial y}\left( \frac{1}{2}\rho V^{2}\right)
\right\vert \propto \left\vert \left( \frac{\partial p}{\partial x}\right)
^{2}\right\vert \text{.}  \label{P3}
\end{equation}%
}

{The behaviours of the equations (\ref{P1}--\ref{P3}) are shown in Fig.5: \
the viscous force term is linear, while the kinematic energy gradient
increases quadratically with the pressure gradient. }Therefore, at low value
of mean velocity $U$, the viscous friction term could balance the
disturbance amplification caused by energy gradient. At large value of $U$,
the viscous friction term may not constrain the disturbance amplification
caused by energy gradient and the flow may transit to turbulence.

For plane Poiseuille flows, the ratio of the energy gradient to the viscous
force term, $K$, is ($\partial p/\partial y=0$),{\ 
\begin{eqnarray}
K &=&\frac{\partial }{\partial y}\left( \frac{1}{2}\rho V^{2}\right) /\left(
\mu \frac{\partial ^{2}u}{\partial y^{2}}\right) =-\frac{2\rho yu_{0}}{h^{2}}%
u_{0}\left( 1-\frac{y^{2}}{h^{2}}\right) /\left( -\mu \frac{2u_{0}}{h^{2}}%
\right)  \nonumber \\
&=&\frac{3}{2}\frac{\rho Uh}{\mu }\frac{y}{h}\left( 1-\frac{y^{2}}{h^{2}}%
\right) =\frac{3}{4}Re\frac{y}{h}\left( 1-\frac{y^{2}}{h^{2}}\right) \text{.}
\label{aa}
\end{eqnarray}%
Here, }$Re=\rho UL/\mu $, and $U=\frac{2}{3}u_{0}$ has been used for plane
Poiseuille flow. $u_{0}$ is the maximum velocity at centerline and $U$ is
the averaged velocity. It can be seen that $K$ is proportional to $Re$ for a
fixed point in the flow field.

The distribution of $u,E,$ and $K$ along the transversal direction for plane
Poiseuille flow is shown in Fig.6. It is clear that there are {\emph{maximum}
of }$K${\ at }${y/h=\pm 0.5774}$, as shown in Fig.6{. This maximum can also
be obtained by differentiating the equation (\ref{aa}) with }$y/h${\ and
letting the derivatives equal to zero. Since we concern the maginitude of
the }$K$ and the velocity profile is symmetrical to the centerline, we refer
the maximum of $K$ as its positive value thereafter. We think that the flow
breakdown of the Poiseuille flow should not suddenly occur in the entire
flow field, but it first takes place at the location of $K_{\max }$ in the
domain, then it spreads out according to the distribution of $K$ value. The
formation of turbulence spot in shear flows may be related to this procedure.

\FRAME{ftbpFU}{3.9877in}{3.5483in}{0pt}{\Qcb{Velocity, energy, and $K$ along
the transversal direction $y/h$ for plane Poiseuille flow, which are
normalized by the their maximum. }}{}{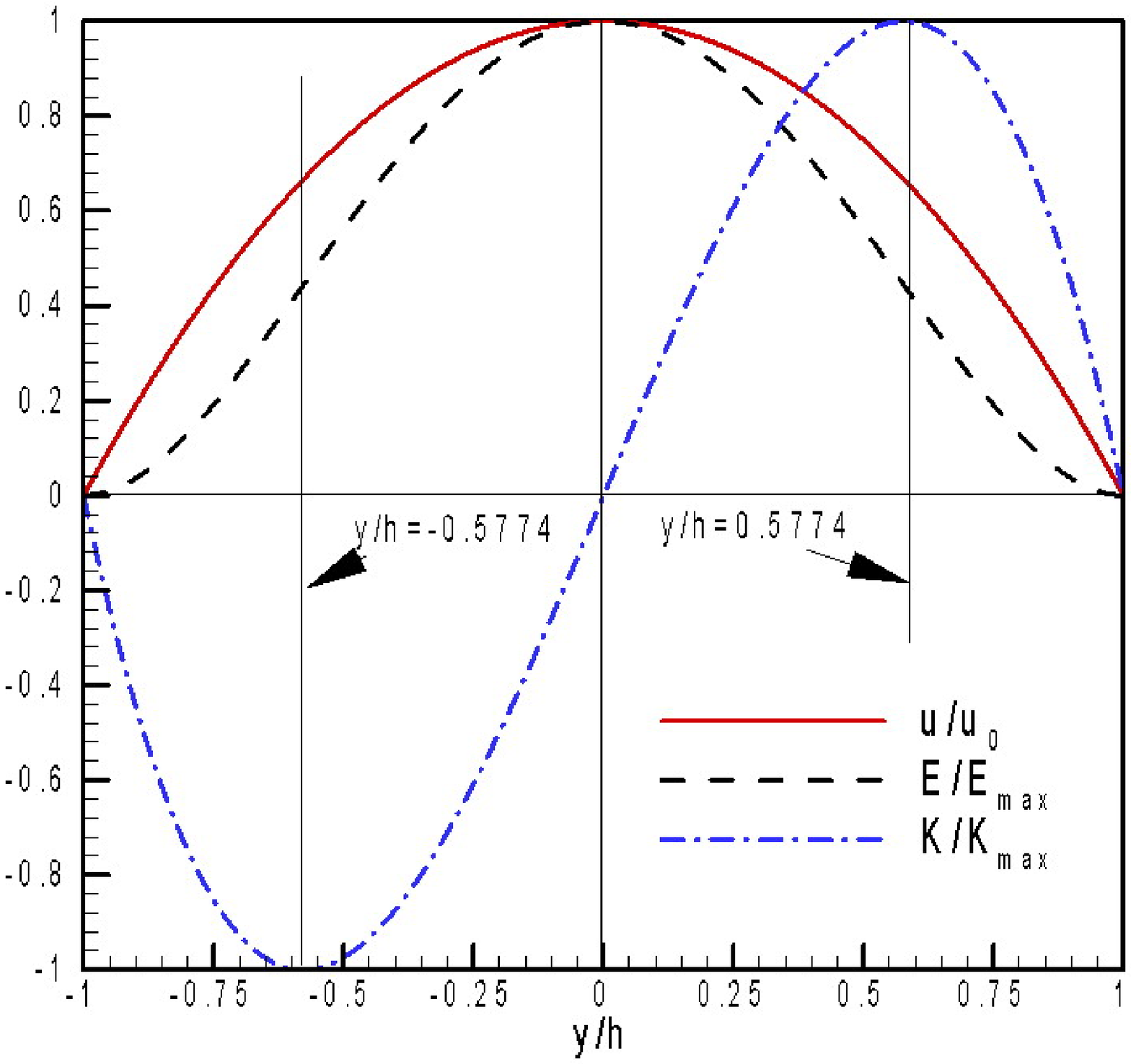}{\raisebox{-3.5483in}{\includegraphics[height=3.5483in]{pp.ps}}}

\FRAME{ftbpFU}{3.9885in}{3.5483in}{0pt}{\Qcb{Velocity, energy, and $K$ along
the transversal direction $r/R$ for pipe Poiseuille flow, which are
normalized by the their maximum. }}{}{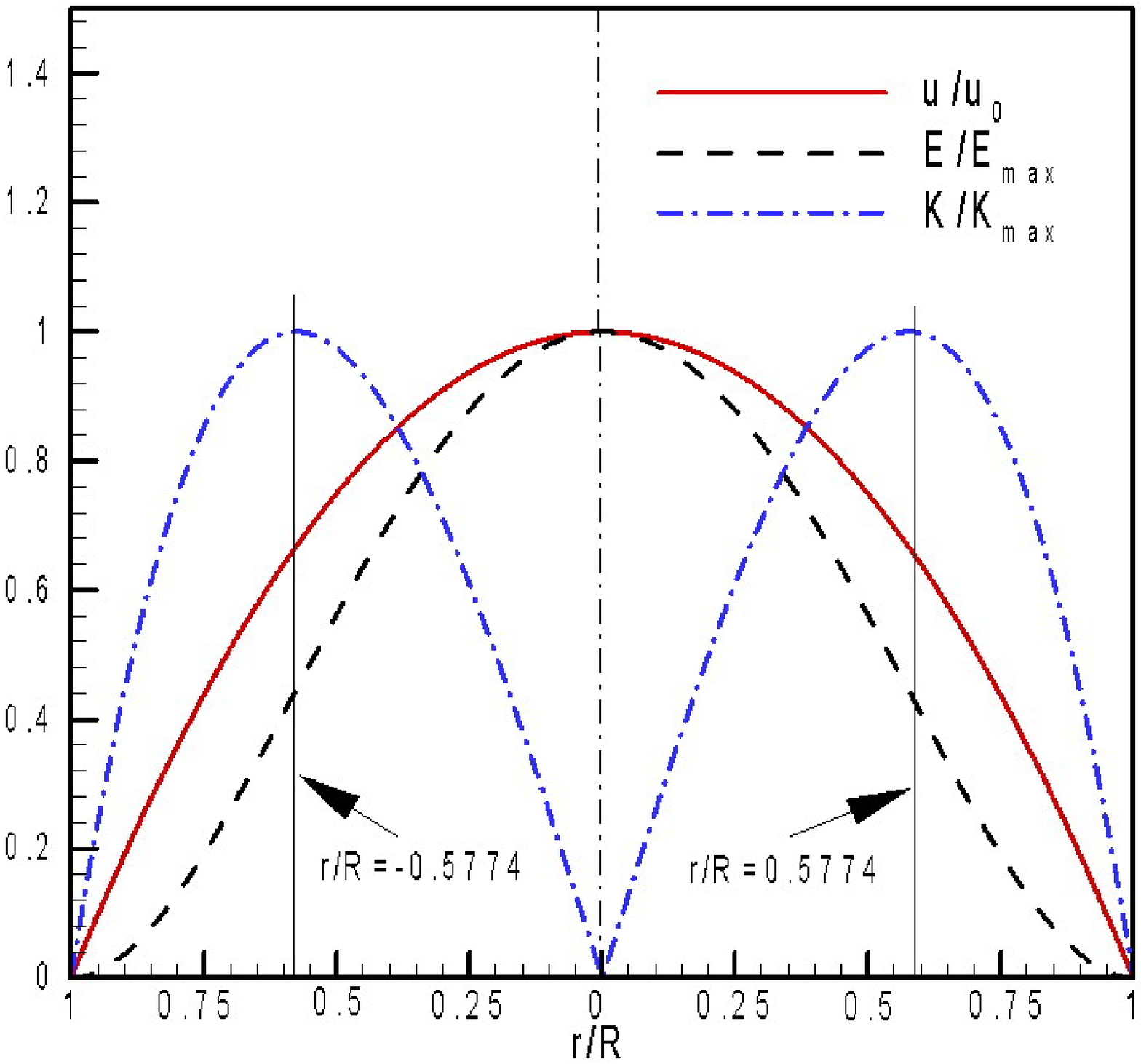}{\raisebox{-3.5483in}{\includegraphics[height=3.5483in]{pp3.ps}}}

\subsection{Pipe Poiseuille Flow}

Similar analysis to the plane Poiseuille flow can be carried out for the
circular Poiseuille flow. For circular pipe Poiseuille flow, the momentum
equation is written as, {\ 
\begin{equation}
0=-\rho \frac{\partial p}{\partial z}+\mu \left( \frac{\partial ^{2}u_{z}}{%
\partial r^{2}}+\frac{1}{r}\frac{\partial u_{z}}{\partial r}\right) \text{,}
\end{equation}%
showing that viscous force term is also proportional to the streamwise
pressure gradient. The }axial velocity is expressed as by integration on
above equation,{\ 
\begin{equation}
u_{z}=u_{0}\left( 1-\frac{r^{2}}{R^{2}}\right) \text{,}
\end{equation}%
where }$u_{0}=-\frac{1}{4\mu }R^{2}\frac{\partial p}{\partial z}$, is the
centerline velocity, $z$ is in axial direction and $r$ is in radial
direction of the cylindrical coordinates, and $R${\ is the radius of the
pipe. }T{he energy gradient can be expressed as below for any position in
the flow field }(noticing $u_{r}=0$){, 
\begin{equation}
\frac{\partial }{\partial r}\left( \frac{1}{2}\rho V^{2}\right) =-\frac{\rho 
}{8\mu ^{2}}R^{2}r\left( 1-\frac{r^{2}}{R^{2}}\right) \left( \frac{\partial p%
}{\partial z}\right) ^{2}\text{.}
\end{equation}%
}

{\ Similar to plane Poiseuille flow, the viscous force term is linear, while
the kinematic energy gradient increases quadratically with the pressure
gradient.}

For pipe Poiseuille flows, the ratio of the energy gradient to the viscous
force term, $K$, is ($\partial p/\partial r=0$),{\ 
\begin{eqnarray}
K &=&\frac{\partial }{\partial r}\left( \frac{1}{2}\rho V^{2}\right) /\mu
\left( \frac{\partial ^{2}u_{z}}{\partial r^{2}}+\frac{1}{r}\frac{\partial
u_{z}}{\partial r}\right) =-\frac{2\rho ru_{0}}{R^{2}}u_{0}\left( 1-\frac{%
r^{2}}{R^{2}}\right) /\left( -\mu \frac{4u_{0}}{R^{2}}\right)  \nonumber \\
&=&\frac{\rho UR}{\mu }\frac{r}{R}\left( 1-\frac{r^{2}}{R^{2}}\right) =\frac{%
1}{2}Re\frac{r}{R}\left( 1-\frac{r^{2}}{R^{2}}\right) \text{.}  \label{bb}
\end{eqnarray}%
Here, }$Re=\rho UD/\mu $, and $U=\frac{1}{2}u_{0}$ has been used for pipe
Poiseuille flow. $u_{0}$ is the maximum velocity at centerline and $U$ is
the averaged velocity.

The distribution of $K$ along the transversal direction for pipe Poiseuille
flow is the same as that for plane Poiseuille flow if it is normalized by
its maximum and $y/h$ is replaced by $r/R$, and the {\emph{maximum} of }$K${%
\ also occurs at }${r/R=0.5774}$, as shown in Fig.7{. }

\section{Comparison with Experiments}

\begin{table}[tbp] \centering%
\begin{tabular}{|l|l|l|l|}
\hline
Flow type & Authors & $Re_{c}$ & $Re_{c}$ \\ \hline\hline
Poiseuille pipe &  &  & $Re=\rho UD/\mu $ \\ \hline
& Reynolds (1883) &  & $2300$ \\ \hline
& Petal \& Head (1969) &  & $2000$ \\ \hline
& Darbyshire \& Mullin(1995) &  & $1760\thicksim 2300$ \\ \hline
& Most literature cited &  & $2000$ \\ \hline\hline
Poiseuille plane &  & $Re=\rho UL/\mu $ & $Re=\rho u_{0}h/\mu $ \\ \hline
& Davies \& White (1928) & $1440$ & $1080$ \\ \hline
& Patel \& Head (1969) & $1350$ & $1012$ \\ \hline
& Carlson et al (1982) & $1340$ & $1005$ \\ \hline
& Alavyoon et al (1986) & $1466$ & $1100$ \\ \hline
& Most literature cited &  & $1000$ \\ \hline
\end{tabular}
\caption{Collected experimental data of the critical Reynolds number 
for plane Poiseuille flow and pipe Poiseuille flow. U is the averaged velocity. 
$u_0$=2U, and D is the diameter of the pipe for pipe Poiseuille flow. 
$u_0$=1.5U, L =2h, and L is the hight of the channel for plane Poiseuille flow.}%
\end{table}%

Experiments for Poiseuille flows indicated that when the Reynolds number is
below a critical value, the flow is laminar regardless of the disturbances.
For circular Poiseuille flow (Hagen-Poiseuille), Reynolds (1883) \cite%
{reynolds} carried out the first systematic experiment on the flow
transition and found that the critical Reynolds number for transition to
turbulence is about $Re_{c}=\rho UD/\mu =2300$, where the $U$ is the
averaged velocity and $D$ is the diameter of the pipe. Now, the most
accepted critical value is $Re_{c}=\rho UD/\mu =2000$ which is demonstrated
by numerous experiments \cite{schlichting}. All the collected data could be
put in a range of $1760$ to $2300$ \cite{darby}. There are also experimental
data for the transition for plane Poiseuille flows in the literature. Davies
and White \cite{davies} showed that the critical Reynolds number for
transition to turbulence is $Re_{c}=\rho UL/\mu =1440$ for plane Poiseuille
flow, where the $U$ is the averaged velocity and $L=2h$ is the width of the
channel. Patel and Head \cite{patel69} obtained a critical value for
turbulence transition, $Re_{c}=2000$ for pipe Poiseuille flow, and $%
Re_{c}=1350$ for channel flow through detailed measurements. Carlson et al. 
\cite{carlson82} found the transition at about $Re_{c}=1340$ for plane
Poiseuille flow using flow visualization technique. Alavyoon et al.'s\ \cite%
{alavyoon} experiments show that the transition to turbulence for plane
Poiseuille flow occurs around $Re_{c}=\rho u_{0}h/\mu =1100$. The most
accepted value of minimum $Re_{c}$ for plane Poiseuille flow is about $%
Re_{c}=\rho u_{0}h/\mu \thickapprox 1000$ \cite{trefethen93} \cite{gross00}.
All the collected experimental data are listed in Table 2. Although these
experiments are done at various different environmental conditions, they are
all near a common accepted value of critical Reynolds number. In the
following, we show that there is a critical value of $K_{\max }$ at which
the flow becomes turbulent. In order to more exactly compare plane
Poiseuille flow to pipe Poiseuille flow at same experimental conditions, we
prefer here to use Patel and Head's data \cite{patel69} to evaluate the
parameters at the critical conditions. Patel and Head's data are also the
best to fit all of the data and are cited by most literature.

Now, we calculate the critical value of $K_{\max }$ at the transition
condition for both plane Poiseuille flow and pipe Poiseuille flow using Eqs.(%
\ref{aa}) and (\ref{bb}), respectively. For plane Poiseuille flow, {\ one
obtains }$K_{\max }{=389}$ a{t the critical Reynolds number }${Re}_{c}{=1350}
$. {For pipe Poiseuille flow, one obtains }$K_{\max }{=385}$ a{t the
critical Reynolds number }${Re}_{c}{=2000}$. These results are shown in
Table 3. In this table, the critical Reynolds number obtained from energy
method is also listed. From the comparison of critical values of $K_{\max }$
for plane Poiseuille flow and pipe Poiseuille flow, we find that although
the critical Reynolds number is different for the two flows, the turbulence
transition takes place at the same $K_{\max }$ value, about $385\backsim 389$%
. This demonstrated that $K_{\max }$ is really a dominating parameter for
the transition, and $K_{\max }$ is a better expression than the $Re$ number
for the transition condition. We can further conclude that energy gradient
theory is better than the linear stability theory for the prediction of
critical Reynolds number of subcritical transition. In this way, the
proposed idea is verified for wall bounded parallel shear flows. Therefore,
it may be presumed that the transition of turbulence in other complicated
shear flows would also depend on the $K_{\max }$ in the flow field.

\begin{table}[tbp] \centering%
\begin{tabular}{|l|l|l|l|l|l|}
\hline
Flow type & $Re$ expression & Eigenvalue analysis & Energy & Experiments & $%
K_{\max }$ at Exp \\ 
&  & $Re_{c}$ & method $Re_{c}$ & $Re_{c}$ & $Re_{c}$ value \\ \hline
Poiseuille pipe & $Re=\rho UD/\mu $ & stable for all $Re$\cite{gross00} & 
81.5 & $2000$\cite{patel69} & $385$ \\ \hline
Poiseuille plane & $Re=\rho UL/\mu $ & $7696$\cite{orszag71} & 68.7 & $1350$%
\cite{patel69} & $389$ \\ \cline{2-6}
& $Re=\rho u_{0}h/\mu $ & $5772$\cite{orszag71} & 49.6 & $1012$\cite{patel69}
& $389$ \\ \hline
Plane Couette & $Re=\rho u_{h}h/\mu $ & stable for all $Re$\cite{gross00} & 
20.7 & $370$\cite{dav, till} & $370$ \\ \hline
\end{tabular}

\caption{Comparison of the critical Reynolds number and the ratio of the 
energy gradient to the viscous force term, $K_{\max }$
for plane Poiseuille flow and pipe Poiseuille flow. The critical Reynolds 
number by energy method is taken from Schmid and Henningson (2001). }%
\end{table}%

Nishioka et al (1975)'s famous experiments \cite{Nishioak75} for plane
Poiseuille flow showed details of the outline and process of the flow
breakdown. The measured instantaneous velocity distributions suggest that
the break down of the flow is a local phenomenon, at least in its initial
stage. As in Fig.8, the base flow is laminar and the instantaneous
distribution of the velocity breaks at the position $y/h=0.50$ ($T=4$ to $6$%
) to $0.62$ ($T=8$ to $9$) by showing an oscillation of velocity in $%
y/h=0.50\thicksim 0.62$. They show an inflectional velocity in this range of 
$y/h$. This result means that the flow breakdown first occurs in the range
of $y/h=0.50\thicksim 0.62$. This coincides to the prediction of our theory,
i.e., the position of $K_{\max }$ is the most dangerous point which occurs
at $y/h=0.5774$. These results are enough to confirm the theory of
\textquotedblleft energy gradient\textquotedblright\ valid at least for
Poiseuille flows (pressure driving flow).

For pipe flow, in a recent study \cite{wedin}, Wedin and Kerswell showed
that there is the presence of the \textquotedblleft
shoulder\textquotedblright\ in the velocity profile at about $r/R=0.6$ from
their solution of travelling waves. They suggested that this corresponds to
where the fast streaks of traveling waves reach from the wall. It\ can be
construed that this kind of velocity profile as obtained by simulation is
similar to that of Nishioka et al's experiments for channel flows. The
location of the \textquotedblleft shoulder\textquotedblright\ is about same
as that for $K_{\max }$. According to the present theory, this
\textquotedblleft shoulder\textquotedblright\ may then be intricately
related to the distribution of energy gradient. The solution of traveling
waves has been confirmed by experiments more recently \cite{hof}.

\FRAME{ftbpFU}{3.9972in}{3.1704in}{0pt}{\Qcb{Instantaneous velocity
distributions in a plane Poiseuille flow (Nishioka et al. 1975) \protect\cite%
{Nishioak75}. Time T corresponding to each distribution is noted on the
trace of the u fluctuation at y/h=0.6, sketched in the figure. Solid circle
is the mean velocity. Uc is the velocity on the channel center-plane. y/h=0
is at the center-plane and y/h=1 is at the wall. Instantaneous velocity U+u
at a point is composed of the mean velocity U and the fluctuation velocity
u. (Courtesy of Nishioka; Use permission by Cambridge University Press).}}{}{%
nishioka1975.ps}{\raisebox{-3.1704in}{\includegraphics[height=3.1704in]{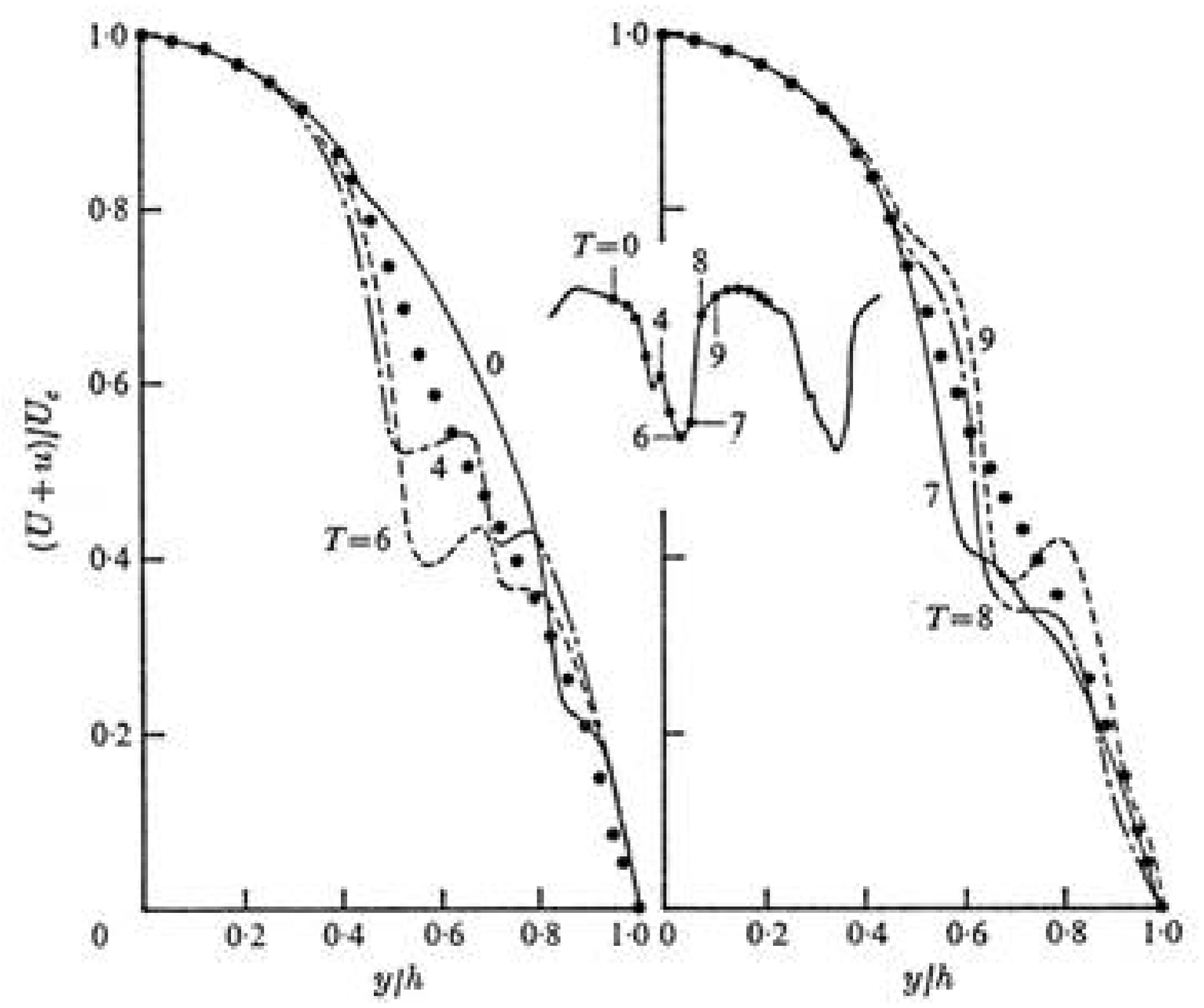}}}

The energy gradient theory can also be used to explain the reason why it is
not appropriate to scale the outer flow or overlap profiles of channel flow
and those of pipe flow at the same Reynolds number turbulent flows\cite%
{wosnik00}.\ \ This is easily understood by the fact that pipe Poiseuille
flow has a same velocity and energy gradient distributions in the radial
direction as the plane Poiseuille flow has in the $y$ direction, but the
former has a smaller hydraulic diameter than the latter and has more viscous
friction. Therefore, the scaling should be carried out at the same $K_{\max
} $ value, but not at the same Reynolds number (Fig.8). \ At a same Reynolds
number, for example, say, $1000$, the plane Poiseuille flow reaches the
critical $Re$ number for transition, while pipe Poiseuille flow is far from
the critical $Re$ number, as shown in Fig.9a. The flow state at these two
flows are definitely different at this Re number. This principle also
applies to turbulence flow range. If we compare the two type of flows at
same $K_{\max }$ value, they should have the same flow behaviour (Fig.9b).

\FRAME{ftbpFU}{4.1165in}{3.0943in}{0pt}{\Qcb{Scaling of plane Poiseuille
flow with pipe Poiseuille flow. (a) Keeping the $Re$ is constant. (b)
Keeping the $K_{\max }$ is constant. \ \ The correlation should be carried
out at the same $K_{\max }$ vaue (b), rather at the same Re (a).}}{}{%
rere1.ps}{\raisebox{-3.0943in}{\includegraphics[height=3.0943in]{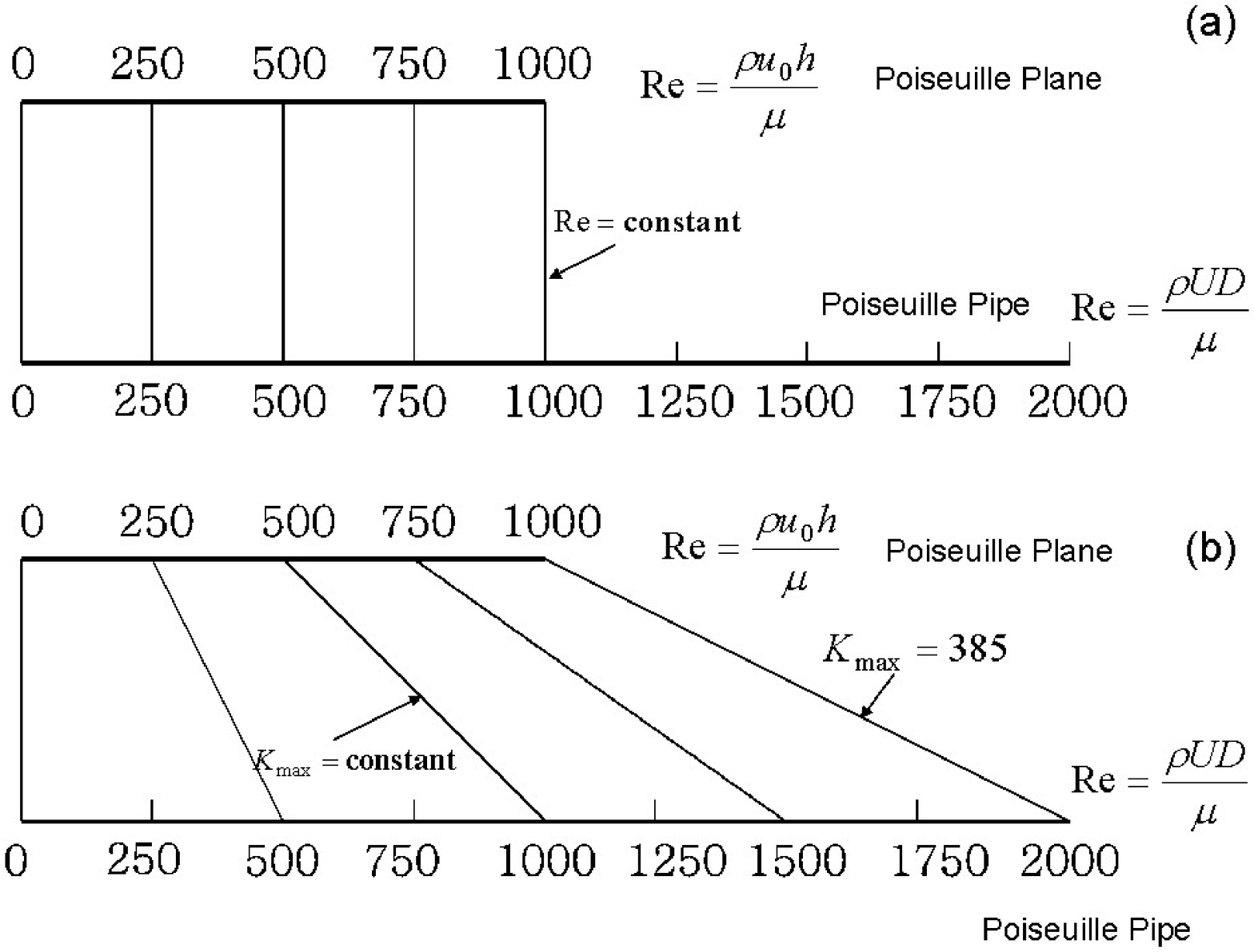}}}

In a separating paper \cite{dou2004} (owing to the space limit here), we
apply the energy gradient theory to the shear driving flows, and show that
this theory is also correct for plane Couette flow. We obtain $K_{\max }=370$
at the critical transition condition determined by experiments below which
no turbulence occurs (see Table 3). This value is near the value for
Poiseuille flows, $385\backsim 389$. The minute difference in the number is
not important because there is some difference in the determination of the
critical condition. For example, the judgement of transition is from the
chart of drag coefficient in Patel and Head \cite{patel69}, while
visualization method is used in \cite{dav, till}. These results demonstrate
that the critical value of $K_{\max }$ at subcritical transition for wall
bounded parallel flows including both pressure driven and shear driven flows
is about $370\backsim 389$.

\FRAME{ftbpFU}{6.3287in}{5.6299in}{0pt}{\Qcb{Comparison of the theory with
the experimental data for the instability condition of Taylor-Couette flow
between concentric rotating cylinders ($R_{1}$=3.80cm, $R_{2}$=4.035 cm). $%
R_{1}$: radius of the inner cylinder; $R_{2}$: radius of the outer cylinder. 
$\protect\omega _{1}$ and $\protect\omega _{2}$ are the angular velocities
of the inner and outer cylinders, respectively. The critical value of the
energy gradient parameter Kc=139 is determined by the experimental data at $%
\protect\omega _{2}=0$ and $\protect\omega _{1}\neq 0$ (the outer cylinder
is fixed, the inner cylinder is rotating). With $K_{c}$=139, the critical
value of $\protect\omega _{1}/\protect\nu $ versus $\protect\omega _{2}/%
\protect\nu $ is calculated using the energy gradient thoery for $\protect%
\omega _{2}/\protect\nu $ = -2200---900 \protect\cite{douTC}.}}{}{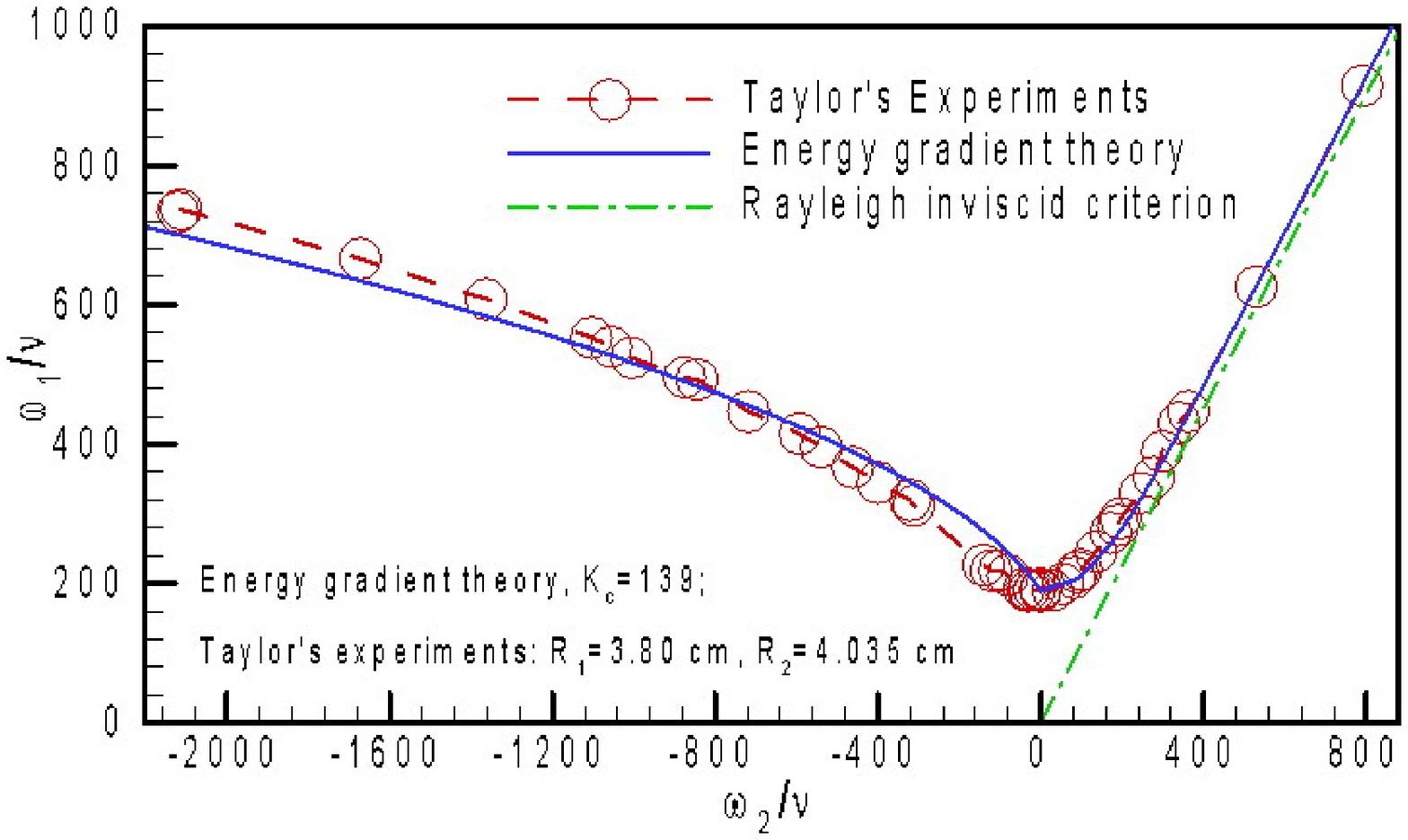}{%
\raisebox{-5.6299in}{\includegraphics[height=5.6299in]{ta44.ps}}}

More recently, the energy gradient theory is applied to the Taylor-Couette
flow between concentric rotating cylinders \cite{douTC}. The detailed
derivation for the calculation of the energy gradient parameter is provided
in the study. The theoretical results for the critical condition of primary
instability obtain very good agreement with Taylor's experiments (1923) \cite%
{Taylor1923} and others, see Fig.10. Taylor (1923) used mathematical theory
and linear stability analysis and showed that linear theory obtained
agreement with the experiments. However, as is well know and discussed
before, linear theory is failed for wall bounded parallel flows. As shown in
this paper, the present theory is valid for all of these said flows.
Therefore, it is concluded that the energy gradient theory is a universal
theory for the flow instability and turbulent transition which is valid for
both pressure and shear driven flows in both parallel flow and rotating flow
configurations.

The Rayleigh-Benard convective instability and the stratified flow
instability could also be considered as being produced by the energy
gradient transverse to the flow (thermal or gravitational energy). The
energy gradient theory can be not only used to predict the generation of
turbulence, but also it may be applied to the area of catastrophic event
predictions, such as weather forecast, earthquakes, landslides, mountain
coast, snow avalanche, motion of mantle, and movement of sand piles in
desert, etc. The breakdown of these mechanical systems can be universally
described in detail using this theory. In a material system, when the
maximum of energy gradient in some direction is greater than a critical
value for given material properties, the system will be unstable. If there
is a disturbance input to this system, the energy gradient may amplify the
disturbance and lead to the system breakdown. This problem will be further
addressed in future study.

\section{Conclusions}

The mechanism for the flow instability and turbulent transition in parallel
shear flows is studied in this paper. The energy gradient theory is proposed
for the flow instability. The theory is applied to plane channel flow and
Hagen-Poiseuille flow. The main conclusions of this study are as follows:

\begin{enumerate}
\item A mechanism of flow instability and turbulence transition is presented
for parallel shear flows. The theory of energy gradient is proposed to
explain the mechanism of flow instability and turbulence transition. It is
stated that the energy gradient in transverse direction tries to amplify\
the small disturbance, while viscous friction in streamwise direction could
resist or absorb this small disturbance. Initiation of instability depends
upon the two roles for given initial disturbance. Viscosity mainly plays a
stable role to the initiation of flow instability by affecting the base flow.

\item A universal criterion for the flow instability initiation has been
formulated for wall shear flows. A new dimensionless parameter
characterizing the flow instability, $K$, which is defined as the ratio of
the energy gradients in transverse direction and that in streamwise
direction, is proposed for wall bounded shear flows. The most dangerous
position in the flow field can be represented by the maximum of $K$. The
initiation of flow breakdown should take place at this position first. This
idea is confirmed by Nishioka et al.'s experiments.

\item The concept of energy angle is proposed for flow instability. This
concept helps to understand the mechanism of viscous instabilities. Using
the concept of energy gradient and energy angle, it is theoretically
demonstrated for the first time that \emph{viscous flow with a velocity
inflection is unstable}.

\item It is demonstrated that there is a critical value of the parameter $%
K_{\max }$ at which the flow transits to turbulence for both plane
Poiseuille flow and pipe Poiseuille flow, below which no turbulence exists. {%
This value is about }$K_{\max }{=385\backsim 389}$. Although the critical
Reynolds number is different for the two flows, the turbulence transition
takes place at the same $K_{\max }$ value.

\item The energy gradient theory is a universal theory for the flow
instability and turbulent transition which is valid for both pressure and
shear driven flows in both parallel flow and rotating flow configurations.
\end{enumerate}

\subsection*{Acknowledgment}

The author would like to thank Professors N Phan-Thien (National University
of Singapore) and JM Floryan (University of West Ontario) for their comments
on the first version of the manuscript.


\begin{thebibliography}{99}
\bibitem{lumly} J.L.Lumley and A.M. Yaglom, A Century of Turbulence, Flow,
Turbulence and Combustion,\textbf{\ 66,} 241-286 (2001).

\bibitem{reynolds} O. Reynolds, An experimental investigation of the
circumstances which determine whether the motion of water shall be direct or
sinuous, and of the law of resistance in parallel channels, Phil. Trans.
Roy. Soc. London A, \textbf{174}, 935-982 (1883).

\bibitem{landau} L.D.Landau and E.M.Lifshitz, \emph{Fluid Mechanics,} 2nd
Ed., (Pergamon, Oxford, 1987), pp.95-191.

\bibitem{drazin} P.G.Drazin and W.H.Reid, \emph{Hydrodynamic Stability},
(Cambridge University Press, Cambridge, 1981), pp.1-250.

\bibitem{schmid01} P.J.Schmid, and D.S.Henningson, Stability and transition
in shear flows, (Springer, New York, 2001).\qquad

\bibitem{trefethen93} L.N.Trefethen, A.E. Trefethen, S.C.Reddy,
T.A.Driscoll, Hydrodynamic stability without eigenvalues, Science, \textbf{%
261}, 578-584 (1993).

\bibitem{gross00} S.Grossmann, The onset of shear flow turbulence. Reviews
of modern physics, \textbf{72}, 603-618 (2000).

\bibitem{wygnanski} I.J.Wygnanski, and F.H. Champagne, On transition in a
pipe. Part 1. The origin of puffs and slugs and the flow in a turbulent
slug. J. Fluid Mech. \textbf{59}, 281-335 (1973).

\bibitem{darby} A.G.Darbyshire and T.Mullin, Transition to turbulence in
constant-mass-flux pipe flow, J. Fluid Mech, \textbf{289,} 83-114 (1995).

\bibitem{orszag71} S.A.Orszag, Accurate solution of the Orr-Sommerfeld
stability equation, J. Fluid Mech, \textbf{50}, 689-703 (1971).

\bibitem{Nishioak75} M. Nishioka, S Iida, and Y.Ichikawa, An experimental
investigation of the stability of plane Poiseuille flow, J. Fluid Mech, 72,
731-751 (1975).

\bibitem{chapamn2002} S.J. Chapman, Subcritical transition in channel flows,
J. Fluid Mech, 451, 35-97 (2002).

\bibitem{orszag80} S.A.Orszag and A.T. Patera, Subcritical transition to
turbulence in plane channel flows, Physical review letters, 45(1980) 989-993.

\bibitem{stuart} J.T.Stuart, Nonlinear Stability Theory, Annual Review of
Fluid Mechanics, \textbf{3,} 347-370 (1971).

\bibitem{LinCC1955} C.-C. Lin, The Theory of Hydrodynamic Stability,
Cambridge Press,1955. Cambridge,1-153.

\bibitem{Betchov} R.Betchov, and W.O.Criminale,Jr., Stability of parallel
flows, Academic Press,New York,1967, 168-170.

\bibitem{joseph76} D.D.Joseph, Stability of fluid motions, Vol.1 and 2,
Berlin : Springer-Verlag , 1976.

\bibitem{hinze} J.O.Hinze, \emph{Turbulence,} 2nd Ed., (McGraw-Hill, New
York, 1975), pp.586-770.

\bibitem{Waleffe} F.Waleffe, Transition in shear flows, nonlinear normality
versus nonnormal linearity, Phys. Fluids, \textbf{7}, 3060--3066 (1995).

\bibitem{baggett} J.S.Baggett, and L.N.Trefethen, Low-dimensionalmodes of
subcritical transition to turbulence, Phys. Fluids, \textbf{9}, 1043-1053
(1997).

\bibitem{Zikanov} O.Y.Zikanov, On the instability of pipe Poiseuille flow,
Phys. Fluids, \textbf{8}, 2923-2932 (1996).

\bibitem{reddy1998} S.C. Reddy, P.J. Schmid, J.S.Baggett, and D.S.
Henningson, On stability of streamwise streaks and transition thresholds in
plane channel flows, J. Fluid Mech, 365, 269-303 (1998).

\bibitem{Meseguer} A.Meseguer, Streak breakdown instability in pipe
Poiseuille flow, Phys. Fluids, \textbf{15}, 1203-1213 (2003).

\bibitem{rayleigh} L. Rayleigh, On the stability or instability of certain
fluid motions, Proc. Lond. Maths. Soc. \textbf{11} 57-70 (1880).

\bibitem{shaqfeh96} E.S.G.Shaqfeh, Purely elastic instabilities in
viscoelastic flows, Annual Review of Fluid Mechanics, \textbf{28,} 129-186
(1996).

\bibitem{dou02a} H.-S.Dou and N. Phan-Thien, Numerical simulation of
viscoelastic flows past a linear array of cylinders by parallel compuation,
Computational Fluid Dynamics 2002, S.Armfield, P.Morgan, and K. Srinivas
Eds., (Springer, Berlin, 2002),323-328.

\bibitem{white} F.M. White, \emph{Viscous Fluid Flow}. (McGraw-Hill, New
York, 2nd Ed., 1991).

\bibitem{schlichting} H.Schlichting and K.Gersten, \emph{Boundary Layer
Theory,} \ (Springer, 8th Ed., Berlin, 2000), pp.415-494.

\bibitem{davies} S.J.Davies and C.M.White, An experimental study of the flow
of water in pipes of rectangular section, Proc. Roy. Soc. A, \textbf{119},
92-107 (1928).

\bibitem{patel69} V.C.Patel, and M.R.Head, Some observations on skin
friction and velocity profiles in full developed pipe and channel flows, J.
Fluid Mech, \textbf{38,} 181-201 (1969).

\bibitem{carlson82} D.R.Carlson, S.E.Widnall, M.F.Peeters, A
flow-visualization study of transition in plane Poiseuille flow, J. Fluid
Mech, \textbf{121,} 487-505 (1982).

\bibitem{alavyoon} F.Alavyoon, D.S.Henningson, P.H.Alfredsson, Turbulent
spots in plane Poiseuille flow--flow visualization, Phys. Fluids, \textbf{29}%
, 1328-1331 (1986).

\bibitem{wedin} H. Wedin, and R.R. Kerswell, Exact coherent structures in
pipe flow: travelling wave solutions, J. Fluid Mech. 508, 333-371 (2004).

\bibitem{hof} B. Hof, C.W. H. van Doorne, J.Westerweel, F.T. M. Nieuwstadt,
H.Faisst, B.Eckhardt, H.Wedin, R.R. Kerswell, F.Waleffe, Experimental
observation of nonlinear traveling waves in turbulent pipe flow, Science,
305 (2004), Issue 5690, 10 September 2004, 1594-1598.

\bibitem{wosnik00} M. Wosnik, L. Castill, W.K. George, A theory for
turbulent pipe and channel flows, J. Fluid Mech, \textbf{421}, 115-145
(2000).

\bibitem{dou2004} H.-S. Dou, B.C.Khoo, K.S.Yeo, and N.Phan-Thien,
Instability of plane couette flow, Technical Report, National University of
Singapore, 2003.

\bibitem{dav} F. Daviaud, J. Hegseth, and P. Berge', Subcritical transition
to turbulence in plane Couette flow, Phys. Rev. Lett. 69, 2511-2514 (1992).

\bibitem{till} N. Tillmark and P. H. Alfredsson, Experiments on transition
in plane Couette flow, J. Fluid Mech. 235, 89 --102 (1992).

\bibitem{douTC} H.-S. Dou, B.C.Khoo, K.S.Yeo, and N.Phan-Thien, Instability
of Taylor-Couette flow between rotating concentric cylinders, submitted to a
Journal, Nov., 2004.

\bibitem{Taylor1923} G. I. Taylor, Stability of a Viscous Liquid Contained
between Two Rotating Cylinders, Philosophical Transactions of the Royal
Society of London. Series A, Vol. 223. (1923), 289-343.
\end{thebibliography}
\end{document}